Modelling the Hyphal Growth of the Wood-decay Fungus *Physisporinus vitreus*


M. J. Fuhr[a,b], M. Schubert[a], F. W. M. R. Schwarze[b] and H. J. Herrmann[a]

[a] *ETH Zurich, Institute for Building Materials, Computational Physics for Engineering Materials, Schafmattstrasse 6, HIF E18, CH-8093 Zurich, Switzerland*

[b] *EMPA, Swiss Federal Laboratories for Materials Science and Technology, Wood Laboratory, Group of Wood Protection and Biotechnology, Lerchenfeldstrasse 5, CH-9014 St Gallen, Switzerland*

*mfuhr@ethz.ch*

*Mark.Schubert@empa.ch*

*Francis.Schwarze@empa.ch*

*hans@ifb.baug.ethz.ch*

Corresponding author: M.J. Fuhr. Phone: +41 44 633 7153

11 figures, 1 table



**Summary**

The white-rot fungus, *Physisporinus vitreus*, degrades the membranes of bordered pits in tracheids and consequently increases the permeability of wood, which is a process that can be used by the wood industry to improve the uptake of wood preservatives and environmentally benign wood modification substances to enhance the use and sustainability of native conifer wood species. To understand and apply this process requires an understanding of how a complex


system (fungus–wood) interacts under defined conditions. We present a three-dimensional fungal growth model (FGM) of the hyphal growth of *P. vitreus* in the heartwood of Norway spruce. The model considers hyphae and nutrients as discrete structures and links the microscopic interactions between fungus and wood (e.g. degradation rate and degree of opening of pits) with macroscopic system properties, such penetration depth of the fungus, biomass and distribution of destroyed pits in early- and latewood. Simulations were compared with experimental data. The growth of *P. vitreus* is characterized by a stepwise capture of the substrate and the effect of this on wood according to different model parameters is discussed.



## 1 Introduction

Hyphal growth and the impact of a fungus are inextricably linked to the underlying substrate and the extracellular digestion of organic matter. The interplay among the chemical composition of the substrate and the concentration of its components, the geometrical structure of the substrate and fungal enzymes determine the growth of fungi. Normally, the natural resources of fungi are physically and chemically complex and structured in space, and often the natural nutrient sources are not continuously uniform, but spatially discrete and heterogeneously distributed. The discrete form of the nutrient source depends on the modelling scale, as well as the porous structure of many natural substrates such as soils or wood.

Wood is a porous material with a complex cellular and hierarchical structure (Fig. 1). On the macroscopic scale, wood consists of growth rings. The three orientations of wood are denoted as longitudinal (along the trunk), radial (from the pith to the bark) and tangential

(tangent to the growth rings). In regions of the world with distinct seasons the growth rings correspond to the stem's annual increase in diameter. On the mesoscopic scale, the growth rings are formed of cells that vary significantly in their geometry and dimensions according to their function. To transport nutrients at the beginning of the vegetative period the tree produces wood with a high porosity (early-wood), whereas latewood is more dense in order to stabilize the tree. At the microscopic level the lumens of the cells (i.e. the voids within the cells) are interconnected through pits. In softwood there are mainly two types of cells: tracheids and rays, which are positioned radially in order to transport nutrients across the growth rings. The pits connecting adjacent tracheids (i.e. bordered pits) have a membrane that controls fluid flow in the tree and most of the bordered pits are located in the radial walls of the tracheids (Bränström 2001). Partial pitting exists on tangential walls (Sirviö 2000, Sirviö & Kärenlampi 1998). The pits connecting tracheids and rays are much smaller (i.e. half-bordered or pinoid pits). Wood is an opaque material, so to analyse its three-dimensional structure, samples are usually sliced into thin sections and analysed by light microscopy (Fuhr et al. 2010, Stührk et al. 2010) or, for non-destructive analysis, X-ray tomography can be used (Van den Bulke et al. 2009). At the submicroscopic level the cell wall is composed of layers of macromolecules, such as cellulose, lignin and hemicelluloses, which are the main nutrient sources of wood-decay fungi, besides water and oxygen. Over time, fungi have evolved different strategies to exploit these nutrients in complex substrates such as wood.

Wood-decay fungi are classified into three types according to their decay pattern: brown rot, white rot (i.e. selective delignification and simultaneous rot) and soft rot (types 1 and 2). An overview of wood decay fungi is presented by Schmidt et al. (1997) and Schwarz & Engels (2004). In the present work we studied the growth of the white-rot fungus, *Physisporinus vitreus*,

in the heartwood of Norway spruce (*Picea abies* (L.) H. Karst.). Generally, in its primary stage of growth this fungus, which is a filamentous basidiomycete, selectively degrades the lignin and hemicelluloses of wood. In the secondary stage it degrades cellulose (Harris et al. 2005). Schwarze and Landmesser (Schwarze et al. 2000) have reported that *P. vitreus* also degrades pit membranes in the heartwood of Norway spruce and that this degradation of pit membranes is pronounced in the primary stage of growth (Schwarze 2007, Schwarze et al. 2008). Therefore, in the initial stage of its growth *P. vitreus* significantly increases the water uptake by wood without significantly reducing the breaking strength of the wood (Hallé 1999, Prosser 1995), but it alters the breaking strength during the secondary stage of its growth. Thus, *P. vitreus* can be use for biotechnological applications, e.g. as shown by Lehringer et al 2009 and Lehringer et al. 2010

The growth and impact of *P. vitreus* have been the subject of many previous studies, although most have dealt with the decay pattern of the fungus (Schwarze & Landmesser 2000, Lehringer et al. 2010) or have tried to quantify the effect of the fungus on macroscopic wood properties such as density (Schwarze & Landmesser 2000), Young's modulus and sound velocity (Schwarze et al. 2008), and permeability (Schwarze et al. 2006, Schwarze 2009). Lehringer et al. (2010) used light microscopy to quantify the patterns of decay of bordered pits, half-bordered pits and cell walls and classified these elements according to fungal activity as 'intact', 'degraded' or 'strongly degraded'. However, physical quantification of the microscopic decay patterns of the fungus is difficult, because of the opacity of wood and the heterogeneity of its structure, so mathematical modelling can potentially be a powerful and efficient method of investigation.

Computational modelling in combination with laboratory experiments can give a deeper insight into the complex interactions between organisms and their environment. For example,

irreversible growth has been extensively investigated in the context of cancer growth, dendritic growth, gelation and penetration in porous media (Araujo et al. 2006, Chopard et al. 1991, Herrmann 1992, Horvath & Herrmann 1991). The irreversible hyphal growth and expansion of *P. vitreus* can be modelled by stochastic processes in both time and space. Wood-decay fungi and their substrates have complex cellular structures from the nanoscopic to the macroscopic level (Fig. 2). Growth of filamentous fungi, especially in such 'multi-scale' materials as soils or wood, is a challenge to quantify and different modelling techniques have been used (Davidson 2007, Prosser 1995). Modelling of filamentous fungi on the colony scale is difficult because the processes governing the growth of the mycelium occur on different scales; for example, uptake of nutrients occurs on the microscopic scale, whereas the transport of nutrients takes place on the meso- or even macroscopic scale. To link the various scales in structured and unstructured soils, Boswell (2008) developed hybrid discrete–continuum models of fungal growth in which hyphae are considered as individual discrete structures and nutrients are modelled as continuous variables. A review of lattice-based and lattice-free models linking hyphal growth to colony dynamics is presented by Boswell and Hopkins (Boswell & Hopkins 2009).

The focus of our investigation was the effect of microscopic parameters, such as the degradation rate, degree of opening of pits and pellet (i.e. starting node) concentration, on macroscopic system properties, such as the penetration depth of the fungus, biomass, distribution of destroyed pits and resultant changes in the wood's permeability. By comparing the macroscopic system properties obtained from simulations with the results of laboratory experiments we hope to improve our understanding on how a complex and difficult to observe system, such as fungus–wood, interacts on the microscopic scale under defined conditions. For this purpose we devised a lattice-free three-dimensional fungal growth model (FGM) that

considers hyphae and nutrients as discrete structures. This FGM is, to our knowledge, the first to simulate hyphal growth and the impact of wood-decay fungi.

## 2 Description of the model

The main focus of the FGM is the identification and investigation of key processes of hyphal growth (e.g. uptake of nutrients) and the resulting increase in wood permeability. In order to investigate these, we developed a model of the microscopic level, focusing on the wood's structure and nutrients (i.e. the substrate) and the tree-like network of the fungus (i.e. the mycelium). Generally, the FGM reduces the enormously complex growth of *P. vitreus* to a pit-to-pit growth of individual hyphae: Starting from an initial state $m = 0$ at each iteration step $m \in \mathbb{N}$ the mycelium may be extended by one edge. This approach uses an adaptive time increment, which enables simulation of the growth of wood-decay fungi from the microscopic (μm) to the macroscopic scale (cm). We used a three-dimensional FGM because of the distinct properties of Norway spruce wood in the longitudinal, radial and tangential directions. The interaction of a fungus with the underlying substrate is enormously complex because of the many feedback mechanisms. The aim of our model is to reduce a complex (biological) system to the essential key processes governing the growth of the fungus and we consider that a model that includes polarization, degradation, transport, branching, growth costs and direction of growth will describe the growth of *P. vitreus* in wood sufficiently for our purpose.

### 2.1 Substrate (Wood)

Norway spruce is a softwood consisting mainly of two cell types. Tracheids are the main cells (90-95%) and the rest are parenchymal cells. During its primary stage of growth *P. vitreus*

degrades the lignified pit membranes of Norway spruce heartwood (Schwarze & Landmesser 2000). The FGM reduces the complex structure of the wood to a network of tracheids (cell walls) connected by bordered pits (nutrient source). Rays and other types of cell wall breaches, such as pinoid pits, are not considered in the model.

The nutrient source is what fungal enzymes act upon. The model assumes that all the essential substances for fungal growth, such as lignin and water, are concentrated in the membranes of bordered pits. The fungus degrades the pit membranes by extracellular digestion and the pit membranes are modelled as points with the attributes

$$R_j \quad j = (1 \ldots N_P),$$

$$F_j^{(m)} \quad \in [0, \nu], \qquad (1)$$

where $R_j$ denotes the coordinates of an arbitrary nutrient point *(j)*, $N_P$ is the number of nutrient points in the system and the variable $F_j^{(m)}$ describes the available amount of nutrients at point *(j)* at iteration step *m*; $\nu$ denotes the initial (*m* = 0) amount of nutrient at point *(j)*. By degrading the lignified pit membrane *P. vitreus* opens the closed connection between two adjacent tracheids, which allows the fungus to exploit the wood and capture the resource. Thereby, the fungus increases the wood's permeability. We assume that the fungus is able to cross the degraded pit membrane and grow into the pits of adjacent tracheids if

$$F_j^{(m)} < \kappa, \qquad (2)$$

where $\kappa \in [0, \nu]$ denotes a specific opening threshold.

Cell walls are the substratum on which the hyphae grow. During the primary growth phase we assume that extracellular enzymes secreted by hyphae do not interact with the cell walls. Hyphal degradation of cell walls commences mainly during the secondary stage. In the present model only the primary stage of hyphal growth is taken into account. Cell walls are



additional boundary conditions within the system: they determine the distance between the pit membranes and limit the accessible set of pit membranes for the fungus.

Figure 3 shows the model wood composed of cell walls and nutrient sources. The tracheids are polyhedrons with six faces. In the model, all tracheids have the same dimension in tangential $w_2$, as well as in the longitudinal $w_3$ directions. In order to model a growth ring, the tracheid width, $w_1$, differs for early-, transition- and latewood. The pits are randomly uniformly distributed along the cell walls according to specific pit densities that depend on the orientation of the underlying cell wall (i.e. tangential or radial pitting) and the position of the cell wall axis within the growth ring (early- and latewood pitting). In order to model overlapping tracheids pits are also located at the tracheid end faces. Differences in size and shape between early- and latewood pits and parenchyma cells (rays) are not considered in this model.

## 2.2 Mycelium (Fungus)

An aggregation of edges and nodes form the mycelium. The edges connect the nodes and represent the filaments, which are the framework of the mycelium. The nodes are at the sites of the pits. Thus,

$$r_i \in R = \bigcup_i R_i$$

$$f_i^{(m)} \geq 0 \qquad\qquad i = \left(1 \dots N_K^{(m)}\right), \qquad (3)$$

$$b_i \geq 0$$

where $N_k^{(m)}$ denotes the number of nodes in the system at iteration step $m$, $r_i$ represents the coordinates of node $(i)$; $R$ is a vector containing the coordinates of the pits; and $f_i^{(m)}$ is the amount of nutrient at node $(i)$. The variable $b_i$ denotes the iteration step at which node $(i)$ is added to the



mycelium. The consumption of nutrients is irreversible. The nodes are connected by edges. The adjacency matrix of the network has the order $(N_k^{(m)} \times N_k^{(m)})$ and is defined as

$$a_{in}^{(m)} \in \{0,1\},$$

$$a_{ii}^{(m)} = 0,$$  (4)

$$a_{in}^{(m)} = a_{ni}^{(m)},$$

where each entry of $a_{in}^{(m)}$ represents a connection between two nodes, *(i)* and *(n)*, receiving the values of 1 and 0 for connected and unconnected nodes, respectively. Thus, the degree $q_i^{(m)}$, the total length $l_i^{(m)}$ of mycelium associated with a node *(i)* and the orientation of a node $c_i^{(m)}$ $\boldsymbol{c_i^{(m)}}$ are given by

$$q_i^{(m)} = \sum_{n=1}^{N_K^{(m)}} a_{in}^{(m)},$$  (5)

$$l_i^{(m)} = 0.5 \sum_{n=1}^{N_K^{(m)}} a_{in}^{(m)} \cdot \|r_i - r_n\|,$$  (6)

$$c_i^{(m)} = \sum_{n=i}^{N_K^{(m)}} \mathcal{H}(b_i - b_n) \cdot a_{in}^{(m)} \cdot \frac{(r_i - r_n)}{\|r_i - r_n\|},$$  (7)

where $\|\cdot\|$ denotes the Euclidian norm. The total hyphal length of the mycelium at iteration step *(m)* is given by summing Eq. 6 over i. $\mathcal{H}(\cdot)$ is the HEAVISIDE step function defined as

$$\mathcal{H}(x) = \begin{cases} 0, & for \quad x < 0 \\ 1, & for \quad x \geq 0 \end{cases}.$$  (8)

There is no restriction concerning the crossing of edges. The model omits autolysis and growth is irreversible. A node with the degree of 1 is called a tip node and nodes with more than one edge are link nodes. Figure 4 shows the network consisting of edges and nodes. The key processes (i.e. polarization of growth, hyphal growth, uptake and concentration of nutrients and branching (lateral and apical)) describe the dynamics of this network and therefore the growing mycelium.

For filamentous fungi the dynamics of the mycelium are dominated by the extension of the filaments at the tips. This apex-located building up of cell walls, called polarization of growth, distinguishes filamentous fungi and is a key aspect of their morphogenesis (Sudbery & Court 2007). The hyphal tip growth is supported by a Spitzenkörper. These sub-apical phase-dark structures found in higher fungi play an important role in the growth and orientation of the hyphal tip (Harris et al. 2005). The Spitzenkörper seems to work as a switching station between the incoming vesicles transporting components for the cell wall and the synthesis of the proteins comprising the cell wall. In higher fungi, hyphal morphogenesis seems to be inextricably linked with the existence of a Spitzenkörper. In order to simulate the dynamics of mycelium we introduced the concept of polarization and Spitzenkörper into the model. The continuous hyphal growth of every polarized fungal cell is modelled by a sequential growth algorithm (i.e. the FGM). Starting from an initial state $m = 0$ at each iteration step $m \in \mathbb{N}$ ONE active node (polarization $p > 0$) is chosen with a probability $P$, using uniformly distributed random numbers, and the mycelium may extended by ONE edge. The probability $P$ of adding an edge to node *(i)* at iteration step $m$ is

$$P_i^{(m)} = \frac{p_i^{(m)}}{\sum_{n=1}^{N_K^{(m)}} p_n^{(m)}}. \tag{9}$$

The polarization $p_i^{(m)}$ of an arbitrary node *(i)* of the mycelium is given by

$$p_i^{(m)} = \mathcal{H}(s_i^{(m)} - 1) \cdot f_i^{(m)}, \tag{10}$$

where $s_i^{(m)}$ is the number of Spitzenkörper at node *(i)*. We define a node as polarized if $p_i > 0$. Spitzenkörper arise from branching and more than one Spitzenkörper per node is possible. The number $s_i^{(m)}$ of Spitzenkörper at node *(i)* is given by

$$s_i^{(m)} = \max\left(\left(s_i^{(m-1)} - 1\right), \left[\frac{f_i^{(m)}}{\beta_i^{(m)}}\right]\right) - \sum_{(n)\neq(i)}^{N_K^{(m)}} a_{in}^{(m)} \cdot \delta_{mb_n}, \qquad (11)$$

$$\beta_i^{(m)} = \begin{cases} \beta_t, & for \quad q_i^{(m)} = 1 \\ \beta_s, & for \quad q_i^{(m)} > 1 \end{cases}, \qquad (12)$$

where $\delta_{ij}$ is the Kronecker delta and $[\cdot]$ is the entire function. The total number of Spitzenkörper in the system at iteration step $m$ is calculated by integrating Eq. 11 over $i$. $\beta_i^{(m)}$ introduces a specific branching threshold that depends on the degree of node *(i)*, i.e. apical ($q_i^{(m)} = 1$) or lateral branching ($q_i^{(m)} > 1$).

Branching is very common in nature, from river networks to living organisms such as fungi (Hallé 1999), which show two types of branching: apical and lateral. If the concentration of nutrients $f_i^{(m)}$ on a link node (lateral) or tip node (apical) exceeds the concentration of $\beta_t \cdot v \cdot \left(s_i^{(m)} + 1\right)$, then branching occurs and the number of Spitzenkörper is incremented by one. $\beta_t$ or $\beta_s$ are specific thresholds for apical ($q_i^{(m)} = 1$) or lateral branching ($q_i^{(m)} > 1$).

As noted before, the FGM is implemented as a sequential algorithm and grows the mycelium by one edge each iteration step $m$. We define the real (laboratory) time increment by

$$dt^{(m)} = \frac{L^{(m)}}{\mu \cdot N_S^{(m)}}, \qquad (13)$$

where $L^{(m)}$ is the length of the edge added to the mycelium at iteration step $m$, $\mu$ is the mean hyphal growth rate, which is equivalent to the averaged velocity of hyphae measured from laboratory experiments, and $N_S^{(m)}$ is the total number of Spitzenkörper in the system. Therefore, the algorithm of the FGM has to run $N_S$ iteration steps to simulate the simultaneous extension of the mycelium at $N_S$ polarized ($p_i > 0$) nodes. If no edge is added to the mycelium (i.e. no event) then the time increment $dt^{(m)} = 0$. The real time $t$ is calculated by integrating Eq. 13 over $m$.



We define the hyphal growth rate as the growth velocity of a single hypha and the mean hyphal growth rate as the averaged growth velocity of several hyphae at a specific region of a colony (e.g. the growth front). The mean hyphal growth rate can be evaluated from laboratory experiments, in our case the radial expansion of a colony on malt extract agar (MEA) at 22°C and pH 5 for different levels of water activity ($a_w$). The $a_w$ is defined as the vapour pressure of a liquid divided by that of pure water at the same temperature. Schubert et al. (2009) found that, apart from nutrients and oxygen supply, pH and $a_w$ play an important role in substrate colonization. On basic MEA medium the optimal conditions for growth of *P. vitreus* are $a_w =$ 0.998, 20°C and pH 5. To reduce the FGM to its essential processes we assume that changes in $a_w$, temperature and pH only affect the mean hyphal growth rate $\mu$. We omit effects of the substrate ( e.g. swelling and shrinking of wood) and assume that the hyphal growth rate is the same in the whole colony from the front to the pellet.

Uptake and concentration of nutrients takes place at the nodes. The evolution equations for the nutrient concentration of a node *(i)* and a pit *(j)* after *m* iteration steps are defined by Eqs 14–19. The terms $\Gamma_i$ and $\Lambda_i$ describe transport processes, while $\Delta_{ij}$ and $\Omega_i$ are the degradation rate and the costs of growth, respectively

$$F_j^{(m)} = F_j^{(m-1)} - \sum_{i=1}^{N_K^{(m)}} \Delta_{ij}^{(m,b_i)}, \tag{14}$$

$$\Delta_{ij}^{(m,b_i)} = \begin{cases} 0, & for \quad m < b_i \\ \alpha_I, & for \quad r_i = R_j, m = b_i \\ \alpha_C \cdot dt^{(m)}, & for \quad r_i = R_j, m > b_i \end{cases} \tag{15}$$

$$f_i^{(m)} = \mathcal{H}(m - b_i - 1) \cdot f_i^{(m-1)} + \sum_{j=1}^{N_P^{(m)}} \Delta_{ij}^{(m,b_i)} + \Gamma_i^{(m,b_i)} - \Lambda_i^{(m)} - \Omega_i^{(m,b_i)}, \tag{16}$$

$$\Gamma_i^{(m,b_i)} = \delta_{mb_i} \left( \sum_{(n,l) \neq (i,i)}^{N_K^{(m)}} a_{in}^{(m)} a_{nl}^{(m)} \frac{f_l^{(m-1)}}{s_l^{(m-1)}+1} + \sum_{(n) \neq (i)}^{N_K^{(m)}} a_{in}^{(m)} \frac{f_n^{(m-1)}}{s_n^{(m-1)}} \right), \tag{17}$$

$$\Lambda_i^{(m)} = \left( \sum_{(n,l) \neq (i,i)}^{N_K^{(m)}} \frac{a_{in}^{(m)} a_{nl}^{(m)}}{s_i^{(m-1)}+1} \delta_{mb_l} + \sum_{(n) \neq (i)}^{N_K^{(m)}} \frac{a_{in}^{(m)}}{s_i^{(m-1)}} \delta_{mb_n} \right) f_i^{(m-1)}, \tag{18}$$

$$\Omega_i^{(m,b_i)} = \delta_{mb_i} \cdot \varepsilon \cdot l_i^{(b_i)}, \tag{19}$$

$\acute{\alpha}_I$ and $\acute{\alpha}_C$ are the initial and continuous degradation rate, respectively, and $\varepsilon$ is the growth costs.

Wood-decay fungi are aerobic organisms producing $CO_2$, water and energy from wood by respiration. They break lignin by oxidoreductase and the degradation of cellulose and hemicelluloses is predominantly by hydrolases (Schmidt 2006). *Physisporinus vitreus* secretes ectoenzymes, such as laccase, to metabolize lignin. In our model the degradation rate $\Delta_{ij}^{(m,bi)}$ represents a more mechanistic view of nutrient uptake by the fungus; that is, if a hypha reaches a pit the transport distance of nutrients to the hyphal tip is very short. Thus, in this initial phase we assume that a hyphal tip accumulates specific amount of nutrients $\alpha_I$ and in a second phase the fungus consumes the nutrients by a constant rate $\acute{\alpha}_C$ (see Eq. 15).

Transport mechanisms are essential for the growth of filamentous fungi and their hyphae in a mycelial network. The mechanisms of nutrient translocation in fungi have not yet been characterized in detail, but a mixture of mass flow, diffusion, cytoplasmic streaming and specific vesicular transport is observed. To encompass these complex biological processes we introduce simple mechanisms into the model, described by $\Gamma_i^{(m,b_i)}$ and $\Lambda_i^{(m)}$ in Eq. 16.

We assume that a specific hyphal length supports a tip node. The supporting length corresponds to the adjacent nodes, as well as to the neighbouring adjacent nodes described by terms 1 and 2 in Eqs 17 and 18. If the mycelium is extended by one edge, a proportion of the

nutrients, depending on the number of Spitzenkörper at the specific node, are transferred to the new tip node.

The direction of hyphal growth depends on the distance and orientation of the node to the reachable pits. For a hyphal node, all pits within the same tracheids, as well as the pits of adjacent tracheids, can be reached if the pit membrane at the underlying hyphal node is open, that is, degraded by the fungus (see Eq. 2). After choosing a node *(i)* according to its polarization every pit *(j)* fulfilling the conditions

$$\cos \Theta \leq \frac{c_i \cdot \left( r_i - r_j \right)}{\|c_i\| \|r_i - r_j\|}$$

and
(20)

$$\xi \leq \|r_i - r_j\|$$

is chosen with the same probability using uniformly distributed random numbers. The model parameters $\xi$ and $\Theta$ describe the direction of growth. At first glance, this approach appears too simple, because of the predominantly forward motion of a hypha (Riquelme et al. 1998). However, wood does in fact canalize hyphal growth, because of the directed distribution of the pits and cell walls, which restricts the available nutrients.

Fungi require nutrients in order to elongate their hyphal cells. In our model the growth costs or nutrients used to synthesize a specific number of hyphal cell walls depend linearly on the length of a new edge (see Eq. 19). The concentration of nutrients $f_n^{(m)}$ in the adjacent node *n* is then reduced by $\Omega_i^{(m,b_i)}$. The mycelium is extended by one edge if $f_n^{(m)}$ of the nutrients in the adjacent node *n* is larger than $\Omega_n^{(m,b_i)}$. If $f_n^{(m)} < \Omega_n^{(m,b_i)}$, growth ceases and no edge will be added to node *n*.

## 3 Simulation

The FGM is a sequential algorithm consisting mainly of three stages. At the beginning, the model is initialized and the boundary conditions are set. The simulation starts at iteration step $m$ = 0 by defining the initial state of the mycelium. Typically, the mycelium simply consists of nodes placed with a density of $n_k^0$ on the surface of the wood specimen. These starting nodes (i.e. the pellets) have a node degree $q_i^0 = 0$, an initial number of Spitzenkörper $s_i^0 = n_s^0$, nutrient concentration $f_i^0 = n_n^0$ and orientation $c_i^0 = n_c^0$. Next, the evolution of the mycelium starts by executing the FGM for as long as the simulation's end is not reached and consists of the following sequential processes. First, a polarized node ($p_i > 0$) of the mycelium is chosen randomly using uniformly distributed random numbers according to Eqs 9 and 10 (polarization). The mycelium is extended by one edge (growth) if pits are available in the vicinity of the chosen node (Eq. 20) and the nutrient source is larger than the growing cost (Eq. 19) of extending the mycelium to the next pit. Next, the real time increment is calculated according to Eq. 13 and all pits underlying a node are degraded according to Eq. 14 (uptake and concentration). Subsequently, all nodes are checked if the condition for branching is fulfilled (branching). Finally, we check if the simulation's end has been reached and if so, the macroscopic variables are estimated (e.g. the distribution of degraded pits).

A typical simulation is shown in Fig. 5 using the model parameters given in Table 1 and one pellet (initial node). We use a rectangular Cartesian coordinate system with the base vectors $e_1$, $e_2$ and $e_3$, where the unit vectors $e_r = e_1$, $e_t = e_2$ and $e_r = e_3$ denote the radial, tangential and longitudinal directions, as shown in Fig. 1. The mean hyphal growth rate $\mu$ was estimated by laboratory experiments Schubert et al 2010. The wood sample consisted of 1,575 tracheids and 96,711 pits. At the beginning of the simulation all pits are closed ($\kappa < \nu$) and have the same

initial amount of nutrient $v$. Initially, the pellet has the attributes $n_s^0 = 1$, $n_n^0 = 3\beta_t/2$ and $n_c^0 = e_1 = [0\ 0\ 1]$. In order to model unlimited extension of the specimen in the radial and tangential directions, periodic boundary conditions are imposed to all faces of the specimen, normal to the given directions. When a hypha passes through such a face, it reappears at the opposite face. The simulation ends when all nutrients are exhausted or a specific number of time steps are reached. In Fig. 5 the simulation was interrupted after 4,000 iteration steps ($\approx 2.5$ day).

## 4 Results and Discussion

The growth and impact of wood-decay fungi is complex because of the heterogeneous structure of the substrate and the mechanism of the growing mycelium of these higher fungi. In the longitudinal direction the tracheids determine the course of the growing mycelium, whereas in the radial direction the rays canalize the hyphae. In the tangential direction the pits are the main pathway for the fungus to move from tracheid to tracheid. In the presented model the rays are not considered in order to reduce the complexity of the substrate. Therefore, the discussion will focus on the penetration of *P. vitreus* in the longitudinal and tangential directions only.

In order to demonstrate the ability of our proposed model to simulate the growth and impact of wood-decay fungi, we discuss the growth pattern, growth characteristics and the impact of the fungus on the substrate and compare the results obtained from our simulations with experimental results. Unless otherwise specified, the parameters given in Table 1 are used throughout.

### 4.1 Growth pattern

Figure 6 shows the modelled growth of *P. vitreus* in the heartwood of Norway spruce after 1,

1.5, 2 and 2.5 days, starting from a single pellet. The wood specimen consisted of 1,516 tracheids

and 96,711 pits. Periodic boundary conditions were applied to surfaces in the tangential and

radial directions.

      The hypha enters a tracheid and grows from pit to pit. The tracheids canalize the growth

of the fungus in the longitudinal direction because of their cylindrical lumens and the initially

closed pits. During growth along the tracheid in a longitudinal direction, the tip of the hypha

splits several times, depending on the apical branching threshold. If the end of a tracheid is

reached, the spread of the mycelium in the longitudinal direction ceases and the mycelium

becomes more dense, because of the uptake of nutrients, and lateral branching occurs. After

some time, the membranes of the bordered pits open ($F_j^{(m)} < \kappa$) and the fungus is able to colonize

an adjacent tracheids, beginning the colonization process again. Therefore, the fungus colonizes

the substrate in the stepwise growth pattern shown in Fig. 6 (d): only a few leading hyphae

penetrate the adjacent tracheids in the longitudinal direction, while the bulk of the mycelium

consists of nodes with coordinates $r_i \cdot e_l < w_3$. Laboratory experiments confirm this characteristic

growth pattern of *P. vitreus* (Stührk et al 2010).

## 4.2 Growth characteristics

In order to analyse the characteristics of growing mycelium in wood we use the same system and

parameters as shown in Fig. 6, but with a different pellet density $n_k^0 = 105$ 1/mm$^2$, totalling 45

pellets. A mean hyphal growth rate of $\mu = 1$ mm/day corresponds to $a_w = 0.972$, 22°C and pH 5

(Herrmann 1992). Figure 7 shows the evolution of the number of Spitzenkörper $N_S$ with time.

After 2 days the mycelium consists of approximately 52,000 nodes and edges and 6,400

Spitzenkörper. After a transient phase, the number of Spitzenkörper increases exponentially (see inset in Fig. 7). Between 0.2 and 1.4 days they fit the law $a \cdot \exp(t/\tau)$ with the parameters $a = 49.8$ and $\tau = 0.38$ and between 1.5 and 2.25 days $a = 188$ and $\tau = 0.56$. Thus, between 0.2 and 1.4 days it takes approximately 1 day to increase the number of Spitzenkörper by a factor of 10. Most Spitzenkörper arise from apical branching and the first lateral branching occurs after approximately 0.5 days. The transient phase is more pronounced, with lateral rather than apical branching. Between 1 and 2.25 days the ratio of the number of Spitzenkörper arising from apical and lateral branching is approximately constant.

For *P. vitreus*, $a_w$ is the most influential factor on the growth rate (Schubert et al 2009). The different phases of growth, with a mean hyphal growth rate $\mu$ varying from 1 ($a_w = 0.972$) to 2 ($a_w = 0.982$) mm/day), are shown in Fig. 8. For the different growth rates the penetration depth and shape of the growth front are similar. During 2 days of growth there are a transient phase and two phases of approximately exponential expansion. The mechanism is the same as the one shown in Fig. 6. In the transient phase the hyphae initiated from the pellets grow without any restrictions, which mean that at every iteration step the mycelium is extended by one edge, but the branching rate is low because of the low nutrient concentration. In this phase all branching is apical. After 12 hours the transient phase ends and the mycelium starts to grow exponentially, as long as the first row of tracheids in the longitudinal direction is colonized. After approximately 36 hours the growth front penetrates adjacent tracheids in the longitudinal direction and reaches a phase of continuous tip production after 1.5 days, again as long as the next row of tracheids in the longitudinal direction is reached. Generally, the penetration of the fungus is characterized by alternating exponential (unrestricted) and restricted growth at a specific frequency. Obviously, this frequency depends on the length of the tracheids in the longitudinal direction.

During unrestricted growth, the total hyphal length and the number of tips of the mycelium increase at the same rate. The ratio between the total length of mycelium and the total number of tips is a constant called the hyphal growth unit (HGU), first postulated by Plomley (1958). The HGU is the average length of a hypha associated with each tip of the mycelium and depends on the fungus and the environmental conditions (e.g. $a_w$ or temperature). In the present model between 0.5 and 1.5 days the HGU is approximately 350 and 550 µm for $\mu = 1.0$ and $\mu = 2.0$, respectively. The smaller the mean hyphal growth rate the denser the mycelium. Figure 9 shows the HGU plotted for different mean hyphal growth rates against time. Between 1 and 2.5 days the HGU oscillates around 350 and 550 µm for $\mu = 1.0$ and $\mu = 2.0$, respectively. After 10 and 28 hours the penetration depth of the growth front increases more than $w_3$·and $2 \cdot w_3$, respectively. Thus, the oscillation occurs because of the stepwise invasion of the substrate in the longitudinal direction. The alternation of restricted and unrestricted growth phases seems to be a characteristic of *P. vitreus*. Obviously, the present model overestimates the fluctuations in the HGU because of the simple substrate used. In reality, the lengths of the tracheids are Gaussian distributed (Brändström 2001) and there are defects in the overlapping regions. In order to analyse the characteristics of the observed fluctuations, further qualitative and quantitative experiments and simulations are necessary.

## 4.3 Impact on early- and latewood

The distribution of pits in Norway spruce wood is inhomogeneous. Because of the different function of the tracheids in early-wood (nutrient transport) and latewood (stability of tree), most of the bordered pits are located in the early-wood tracheids and both the size and number of bordered pits in latewood is smaller. Therefore, the intensity of fungal activity in the first stage



of growth strongly correlates with the underlying substrate, because of the different amounts and availability of nutrients in early- and latewood.

Figure 10 shows the number of open pits plotted against time for different mean hyphal growth rates. The number of open pits increases exponentially with time, and surprisingly, there are no fluctuations resulting from the stepwise growth.

In order to analyse fungal activity in early-, transition- and latewood we use a model specimen with 25, 40 and 100 tracheids/mm, respectively, in the radial direction, with a pellet density of 25 mm (Fig. 5a). Thus, early-, transition- and latewood occupy the same volume. The simulation parameters are the same as shown in Table 1, except for the growth cost $\varepsilon$. Figure 11 shows the mycelium, the distribution of remaining nutrients and the number of open pits (see Eq. 2) for all pits occupied by more than one node after 2.5 days of incubation for $\varepsilon \in [10^{-4}, 10^{-3}]$.

Generally, there is much higher fungal activity in early- and transition wood than in latewood with regard to the amount of mycelium, the growth front and the number of degraded pits. Because of the strong correlation between the number of open pits and the remaining nutrients we focus our analysis on the number of open pits. The difference in the number of open pits in early-, transition-and latewood for $\varepsilon = 10^{-4}$ and $\varepsilon = 10^{-3}$ is approximately 35%, 11% and 43%, respectively. In total, after 2 days approximately 77% and 75% of the occupied pits are open for $\varepsilon = 10^{-4}$ and $\varepsilon = 10^{-3}$, respectively, while the absolute number of open pits is approximately 17% higher in measurement Fig. 11 (c) than in Fig. 11 (d). As anticipated, the cost of growing affects the number of occupied pits and the growth front: the higher the growth cost the smaller the polarity and therefore the velocity of the fungus.

A detailed view of the distribution of open pits in measurement Fig. 11 (c) shows that, surprisingly, there is a significant difference in the number of open pits: after 2 days of



incubation approximately 71%, 91% and 55% of occupied pits are open in early-, transition- and

latewood, respectively. Because of the linear degradation rate $\dot{\alpha}_C$, we would expect

approximately the same proportion of open pits, but the mean number of nodes per pit in

transition wood is higher than in early- and latewood. Thus, the initial degradation rate plays an

important role by opening the pits and therefore enhancing the spread of the mycelium in the

tangential and radial directions. Latewood seems to be less accessible to hyphae than early- and

transition wood.

The lower accessibility of latewood regions by *P. vitreus* was also supposed by Lehringer

et al. 2010, because of the cavities and notches that are mainly in the latewood. Perhaps *P.*

*vitreus* creates its own voids in order to colonize regions of Norway spruce with fewer and

smaller pits. If we compare the distributions between studies, there is a difference: Lehringer et

al. (2010) found no significant differences between the degradation of bordered pit membranes in

early- and latewood whereas in our simulation the number of strongly degraded pits ($f_i^{(m)} < 0.2 \cdot v$)

was higher in early-wood than in latewood (Fig. 11). However, this result depends on the

definition of 'strongly degraded' and the incubation conditions. Thus, further qualitative and

quantitative experiments and simulations are necessary to explain the colonization strategy of the

fungus.

## 5    Conclusion

We present a three-dimensional mathematical model of hyphal growth and the impact of the

white-rot fungus *P. vitreus*. The model focuses on the structure of the wood (i.e. nutrients and

substrate) and the tree-like network of the fungus (i.e. the mycelium), the evolution of which

with time is governed by key processes such as polarization, uptake and transport of nutrients

and the cost of growth. The model reduces the enormously complex growth of *P. vitreus* to a pit-to-pit motion of individual hyphae. Such an approach uses an adaptive time increment, which enables simulation of the growth of *P. vitreus* from the microscopic (μm) to the macroscopic scale (cm).

We found that the fungus captures the substrate in a stepwise pattern, with alternating phases of restricted and unrestricted growth, and we compared the simulation results qualitatively with experimental results (Stührk et al. 2010, Lehringer et al. 2010). We found that fungal activity in early-wood is much higher than in latewood after 2 days of incubation, whereas Lehringer et al. (2010) found no significant differences in fungal activity. The discrepancy in these findings may be explained by the different observation times, but further qualitative and quantitative experiments and simulations are necessary to explain the colonization strategy of the fungus.

In particular, the influence of tracheid length on the penetration velocity of wood-decay fungi in the longitudinal direction is of interest, as well as the impact of the fungus on the substrate (e.g. changes in permeability). In order to simulate the permeability changes induced by *P. vitreus*, a more accurate wood model in terms of pit distribution and pit size, as well as parenchymal cells, is essential. Images from synchrotron radiation tomography show the anatomical features of wood, such as pits and rays, and confocal laser scanning microscopy offers a technique for measuring the evolution and spatial distribution of biomass in wood. These microscopic techniques will enable the development of a more realistic model that includes anastomosis, metabolism and the presence of an inhibitor.

Generally, hyphal growth models are limited up to the centimetre scale because of the growth of millions of hyphal tips in the small specimens of wood. However, we consider that the

present model enables analysis of the effects of microscopic parameters, such as the degradation rate and degree of opening of pits, on macroscopic system properties such as penetration depth of the fungus, biomass and distribution of destroyed pits in early- and latewood. This understanding of how complex systems (e.g. fungus-wood) interact under defined conditions is crucial for the biotechnological applications of *P. vitreus*.

## 6 Acknowledgements


The authors gratefully acknowledge to C. Stührk, T. Pähtz, F. K. Wittel and P. Niemz for helpful discussions and express their gratitude to the Swiss National Science Foundation (SNF) No. 205321-121701 for its financial support. Dr Kerry Brown is thanked for proofreading the manuscript.

**Figure captions**

Figure 1: Cellular structure of wood. Tracheids and rays comprise the cellular structure of Norway spruce (*Picea abies* [L.] H. Karst.). Bordered and pinoid pits connect the cell lumens

(voids within the cells). The cells vary significantly in their geometry and dimensions and form a growth ring consisting of a continuous transition from early-wood (high porosity) and transition-wood to latewood (low porosity). As shown, at the boundary between adjacent growth rings there is a sharp transition from early- to latewood. The three orientations of wood are denoted as longitudinal (L), radial (R) and tangential (T).

Figure 2: Hierarchical structure of wood and fungi. Wood-decay fungi and their substrates have a complex cellular and hierarchical structure from the nanoscopic to the macroscopic scale. Modelling wood-decay fungi is a challenge because of the processes governing the growing mycelium being performed on different scales (e.g. uptake of nutrients on the microscopic scale while transport of nutrients takes place on the macroscopic scale).

Figure 3: Model of Norway spruce wood. The model reduces the complex wood structure to a network of tracheids connected by bordered pits. The pits are randomly uniformly distributed with density $\rho$ along the cell wall axis (dotted line), depending on the orientation of the underlying cell wall (tangential or radial pitting) and the position of the cell wall in the growth ring (early- and latewood pitting). In order to model a growth ring, the tracheid width $w_1$ for early-, transition- and latewood differs.

Figure 4: Model of *Physisporinus vitreus*. An aggregation of edges and nodes forms the mycelium. The position of the nodes is restricted to the pits. Spitzenkörper denote polarized nodes. Branching occurs if the concentration of nutrients at a node exceeds a specific threshold.

Figure 5: Specimens used throughout the study. (a) Typical simulation: penetration of the fungus (black) after 1 day (approximately 100 iteration steps of the fungal growth model) in the longitudinal direction using the parameters shown in Table 1. The wood sample (grey) consists of 1575 tracheids and 96,711 pits. Periodic boundary conditions are applied to the specimen's



faces, normal to the tangential (T) and radial (R) directions. In this case fungal growth started from a single initial node (pellet) with one Spitzenkörper. (b) In order to analyse fungal activity in early-wood (EW), transition-wood (TW) and latewood (LW) we used specimens consisting of 25, 40 and 100 tracheids/mm, respectively, in the radial direction, with a pellet density $n_k{}^0 = 25$ 1/mm$^2$. Thus, EW, TW and LW occupy the same volume.

Figure 6: Evolution of the mycelium. (a)–(d) Growth of the fungus *Physisporinus vitreus* in heartwood of Norway spruce after 1, 1.5, 2 and 2.5 days, starting from a single pellet. The wood specimen consisted of 1516 tracheids and 96,711 pits. Periodic boundary conditions were applied to surfaces in the tangential and radial directions. The model parameters are given in Table 1. After 2 days, only a few leading hyphae have penetrated into adjacent tracheids in the longitudinal direction and the bulk of the mycelium remains in the first section of the specimen (nodes with coordinates $r_i \cdot e_l < w_3$).

Figure 7: The number of Spitzenkörper plotted against time for a mean hyphal growth rate of $\mu = 1.0$ mm/day. Between 0.2 hours and 1.4 days the Spitzenkörper evolution fits an exponential law $a \cdot \exp(t/\tau)$ with the parameters $a = 49.8$ and $\tau = 0.38$. Most Spitzenkörper arise from apical branching. Between 1.4 and 2.25 days the ratio of the number of Spitzenkörper arising from apical and lateral branching is approximately constant.

Figure 8: The number of Spitzenkörper plotted against time for hyphal growth rates varying from $\mu = 1.0$ to $\mu = 2.0$ mm/day. During 2 days of growth there are a transient phase and two phases of approximately exponential expansion. Generally, the penetration of the fungus is characterized by alternating exponential (unrestricted) and restricted growth at a specific frequency.

Figure 9: The hyphal growth unit (HGU) plotted for different hyphal growth rates. The HGU is defined as the ratio of the total length of the mycelium and the total number of tips. For

unrestricted growth, the HGU is a constant. Between 0.5 and 1.5 days the HGU oscillates around 350 and 550 μm for $\mu = 1.0$ and $\mu = 2.0$, respectively. The oscillation occurs because of the stepwise invasion of the substrate by the fungus in the longitudinal direction. The alternation of restricted and unrestricted phases of growth seems to be a characteristic of *Physisporinus vitreus*. After 10 and 28 hours the penetration depth of the growth front increases more than $w_3$ and $2 \cdot w_3$, respectively (arrows).

Figure 10: The number of open pits plotted against time for different mean hyphal growth rates increases exponentially with time. Surprisingly, there are no fluctuations in the evolution of the number of open pits arising from the stepwise growth.

Figure 11: Fungal activity characterized by the mycelium (a, b), the distribution of remaining nutrients (c, d) and the number of open pits (e, f) (see Eq. 2) for all pits occupied by more than one node after 2.5 days of incubation for $\varepsilon \in [10^{-3}, 10^{-4}]$. There is much higher fungal activity in early- and transition wood than in latewood with regard to the amount of mycelium, the growth front and the number of degraded pits.

**Table caption**

Table 1: Parameters used in the development of the fungal growth model. We used a rectangular Cartesian coordinate system with the base vectors $e_1$, $e_2$ and $e_3$, where the unit vectors $e_r = e_1$, $e_t = e_2$ and $e_r = e_3$ denote the radial (R), tangential (T) and longitudinal (L) directions as shown in Fig. 1.



| Substrate (Wood): | | | |
|---|---|---|---|
| tracheid width / height / length | $w_{1,2,3}$ | 0.001-0.04 / 0.04 / 2 | mm |
| pit density early-wood | $\rho_{ET}$ ($\rho_{ER}$) | 0.02 ($\rho_{ET}/2$) | 1/mm |
| pit density latewood | $\rho_{LT}$ ($\rho_{LR}$) | $\rho_{ER}/2$ ($\rho_{ER}/2$) | 1/mm |
| | | | |
| **Mycelium (Fungus):** | | | |
| mean hyphal growth rate | $\mu$ | 1 | mm/d |
| Growth cut-off length | $\theta$ | L/20 | mm |
| Growth cut-off angle | $\xi$ | 90 | ° |
| Growth costs | $\varepsilon$ | $10^{-17}$ | mol/mm |
| Pit inital nutrient | $\nu$ | $4*10^{-13}$ | mol |
| Pit initial degradation rate | $\alpha_I$ | $\nu/5$ | mol |
| pit degradation rate | $\alpha_C$ | $\nu/10$ | mol/d |
| Pit opening | $\kappa$ | 0.5 | mol/mol |
| Apical branching threshold | $\beta_t$ | $2*\nu$ | mol |
| Lateral branching threshold | $\beta_s$ | $\nu/4$ | mol |
| | | | |
| **Simulation:** | | | |
| Initial pellet density | $n_k^0$ | 12 | $1/mm^2$ |
| Initial number of Spitzenkörper | $n_s^0$ | 12 | - |
| Initial nutrient concentration | $n_n^0$ | $\beta_t/2$ | mol |
| Initial orientation | $n_c^0$ | $e_1$ | - |



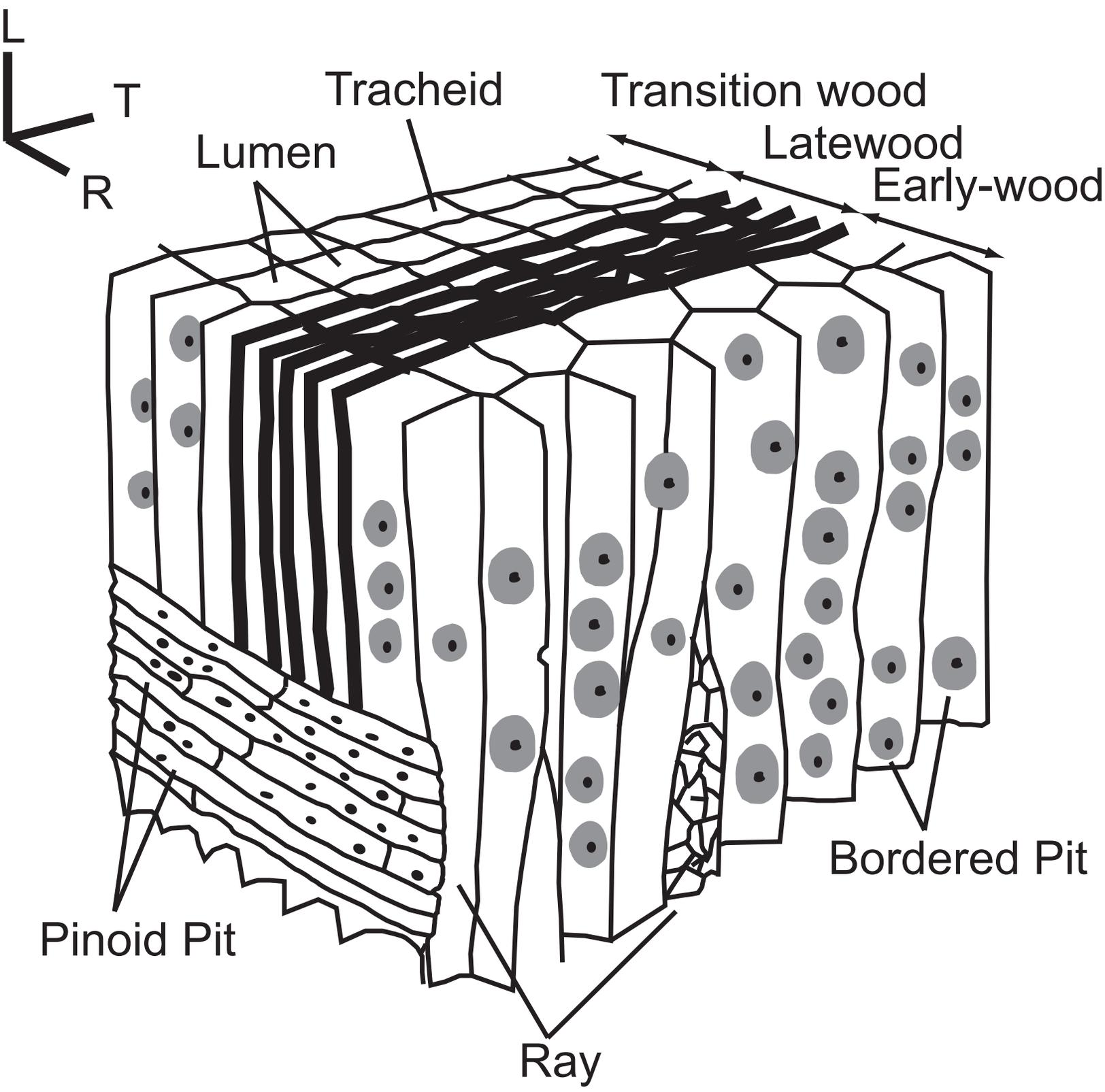



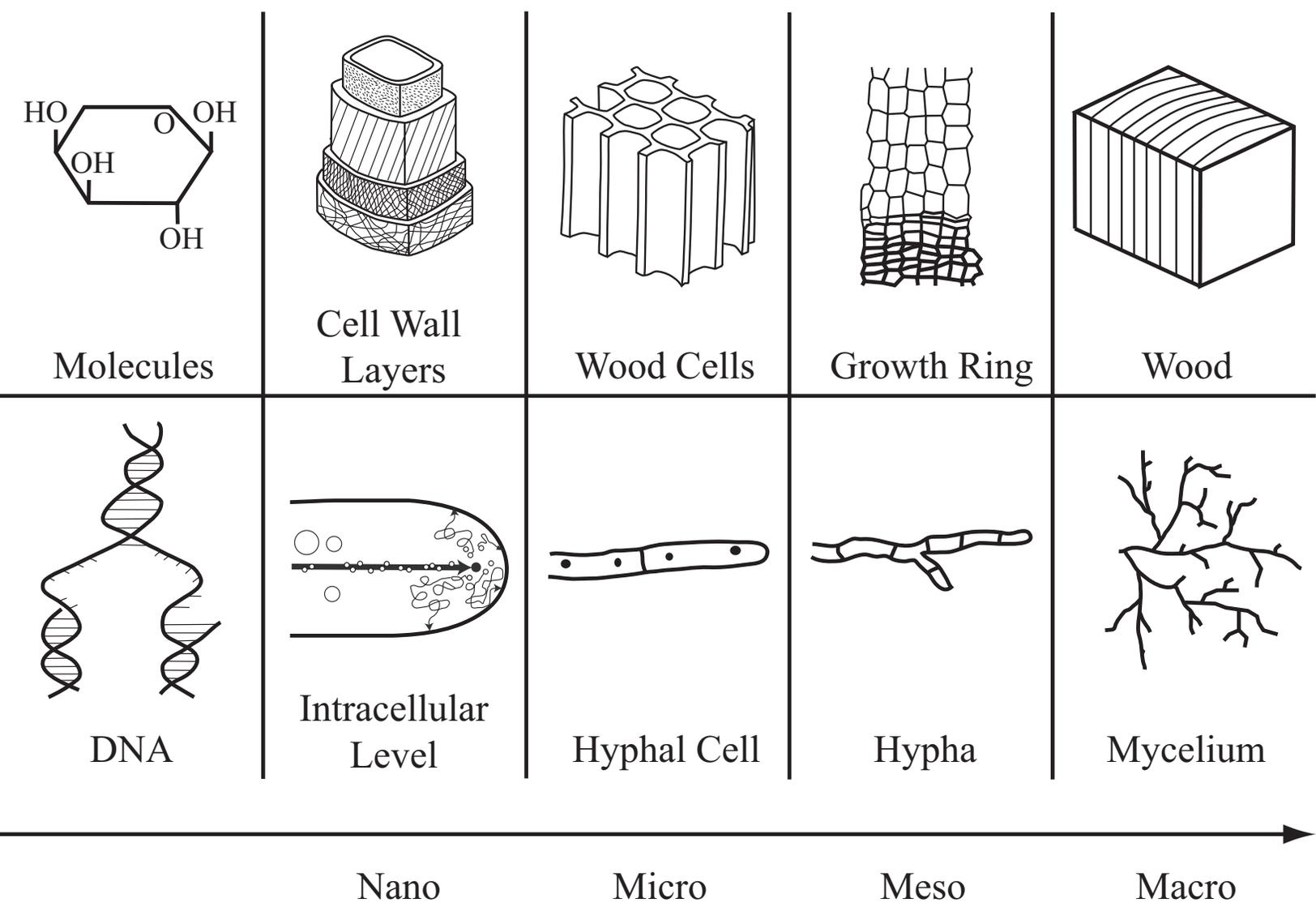

| Molecules | Cell Wall Layers | Wood Cells | Growth Ring | Wood |
| DNA | Intracellular Level | Hyphal Cell | Hypha | Mycelium |

Nano          Micro          Meso          Macro



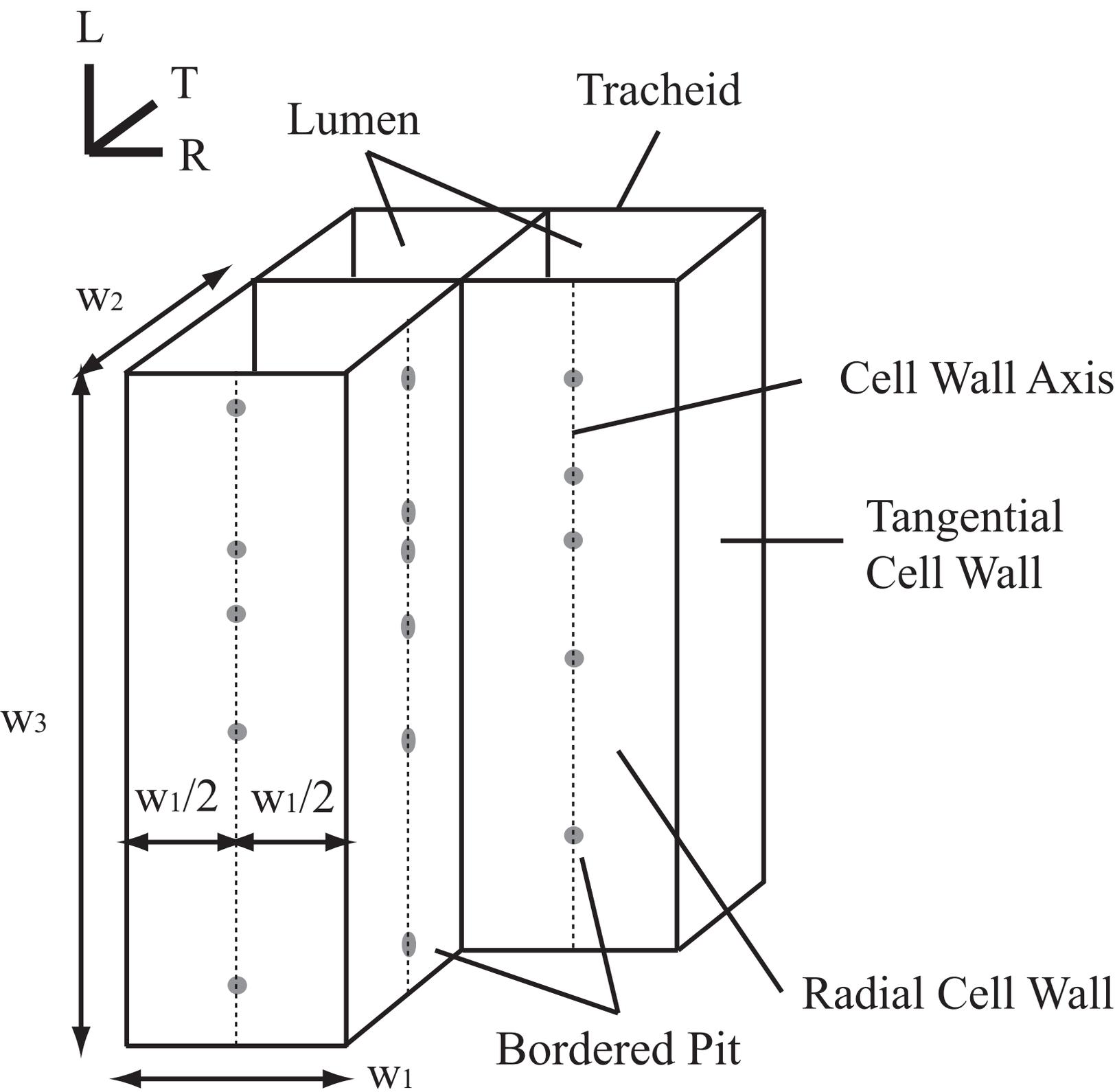

fig4_smallcolumn.eps

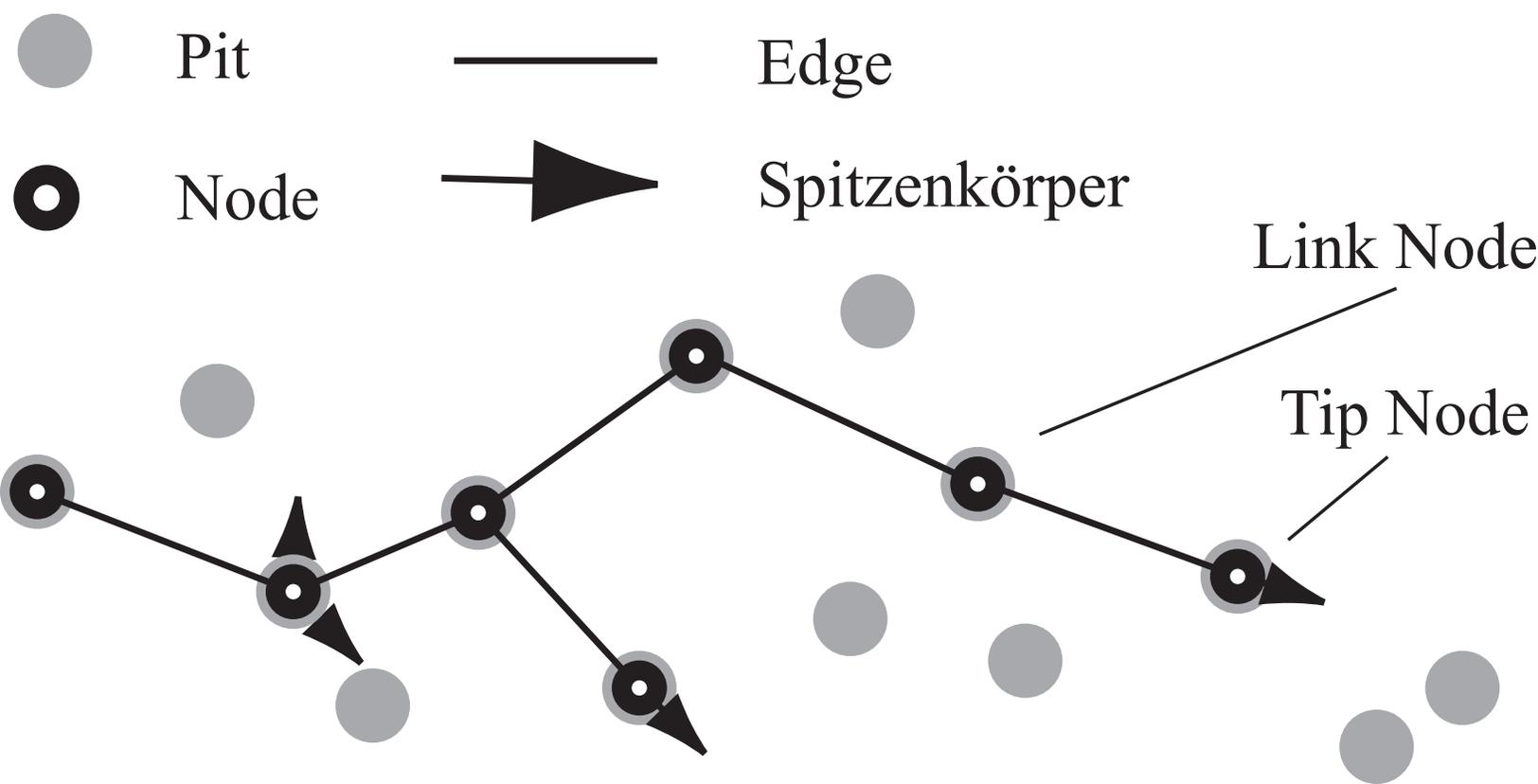

Pit

Node

Edge

Spitzenkörper

Link Node

Tip Node



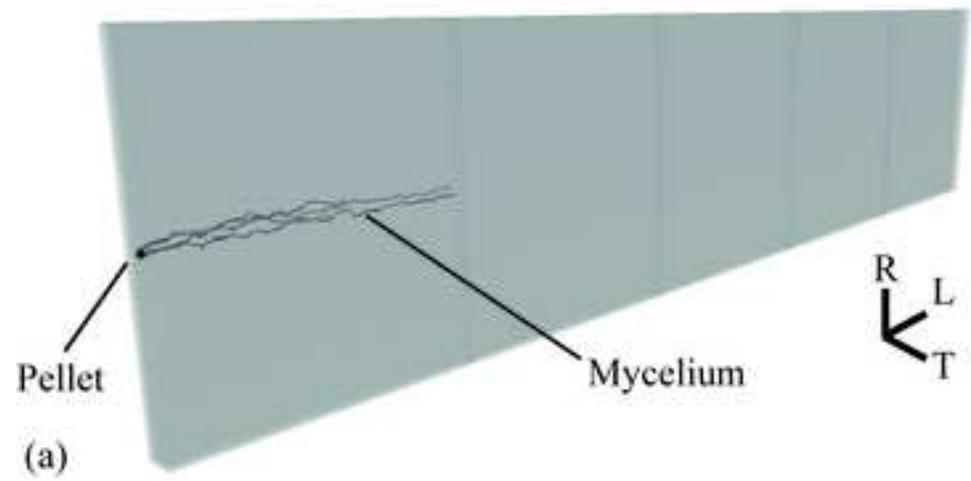

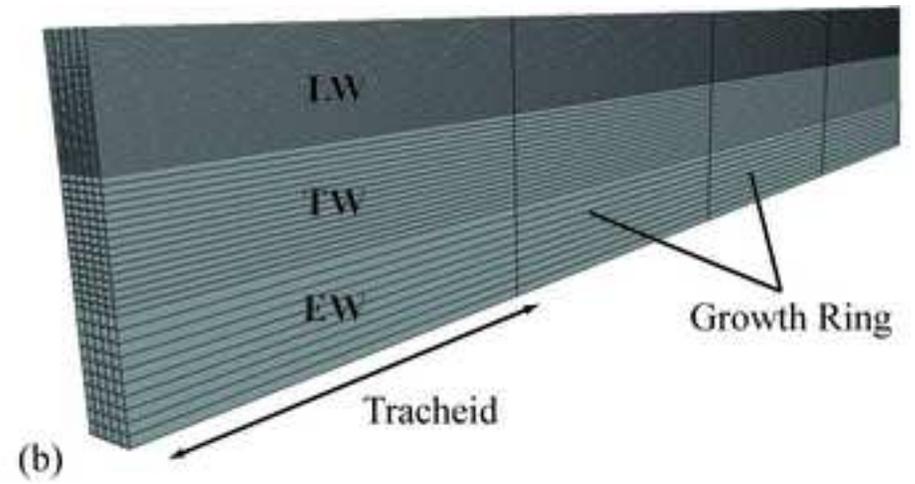

(a) Pellet    Mycelium

R
L
T

(b) LW    TW    EW    Growth Ring    Tracheid



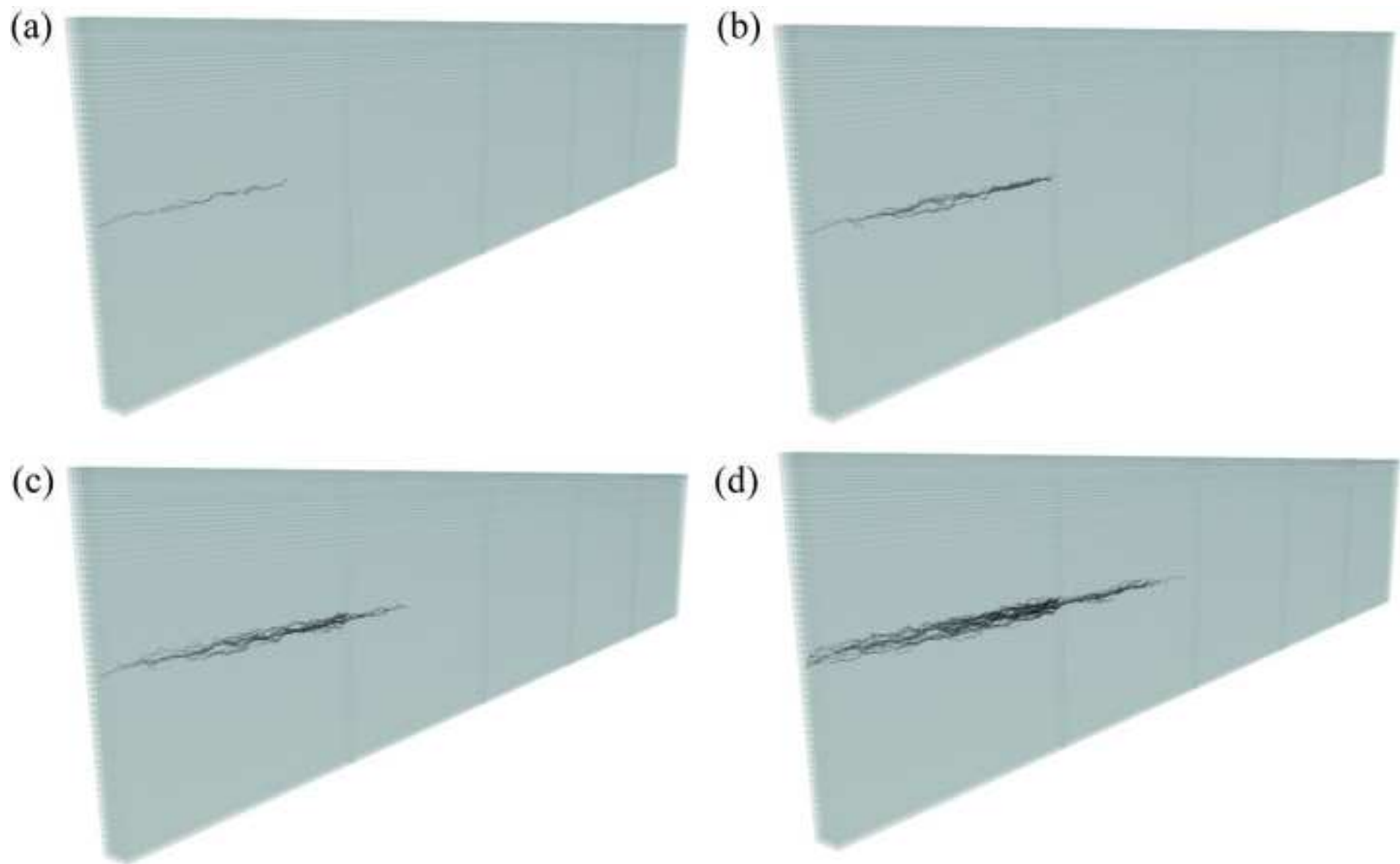



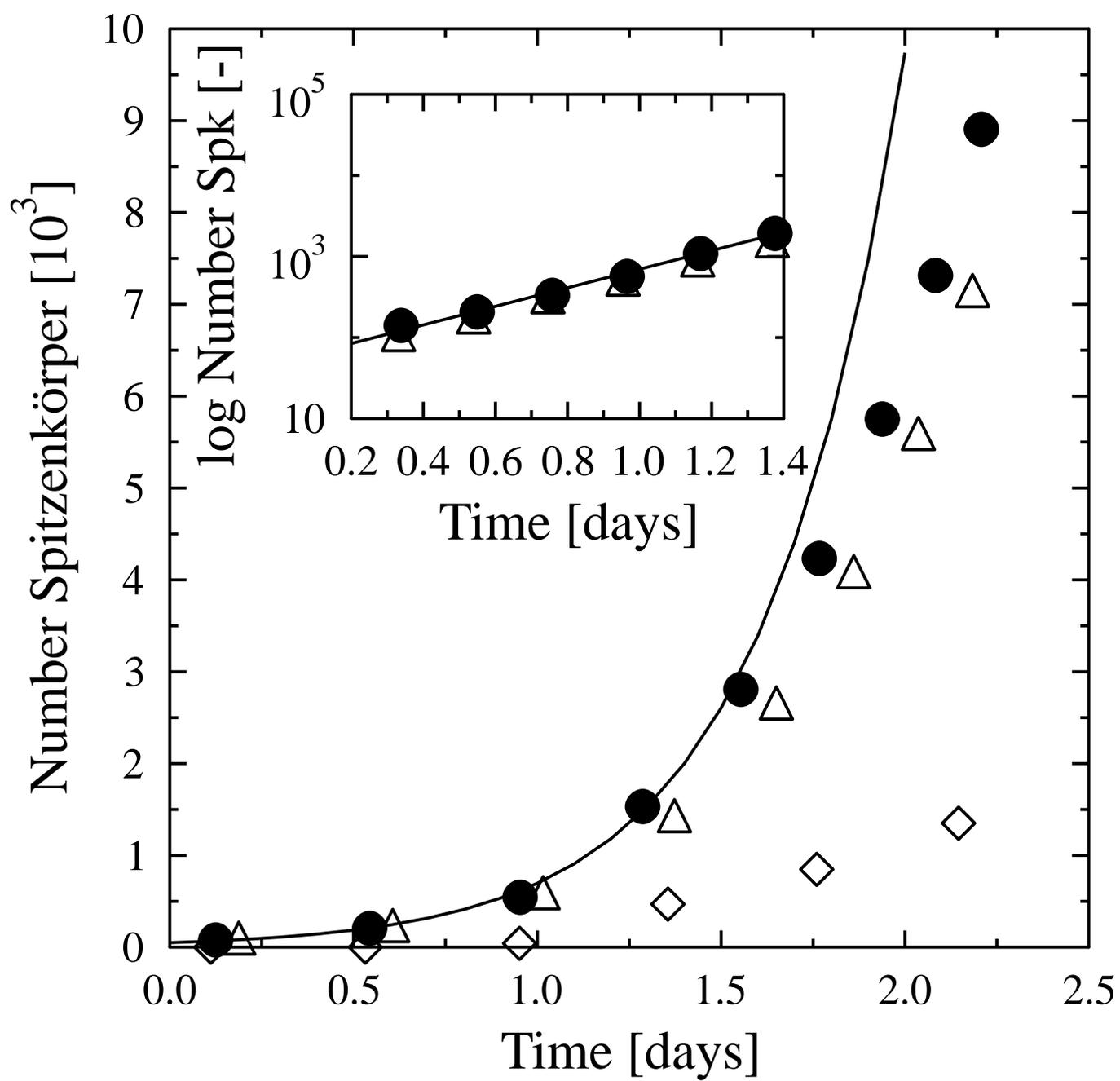



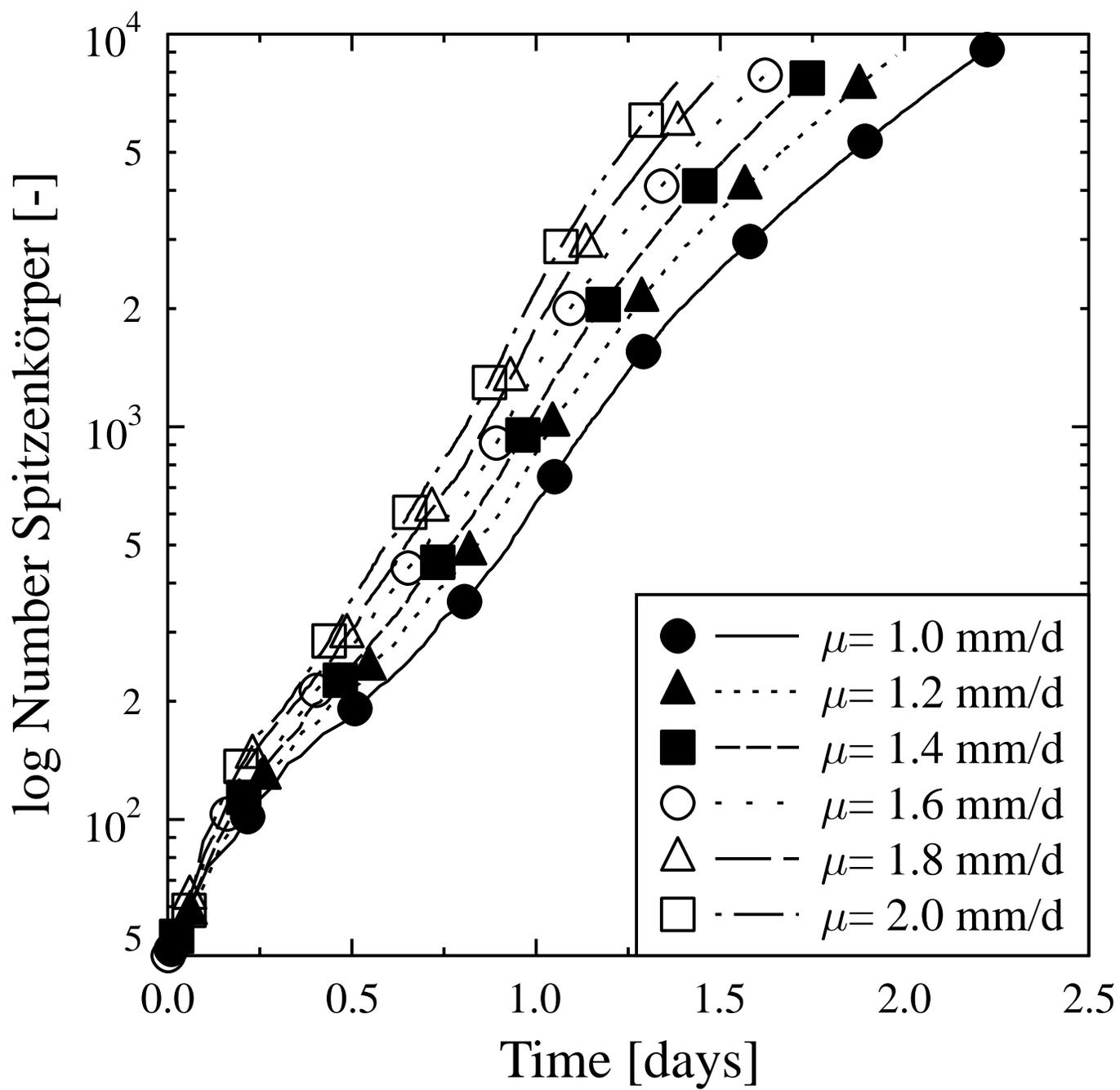



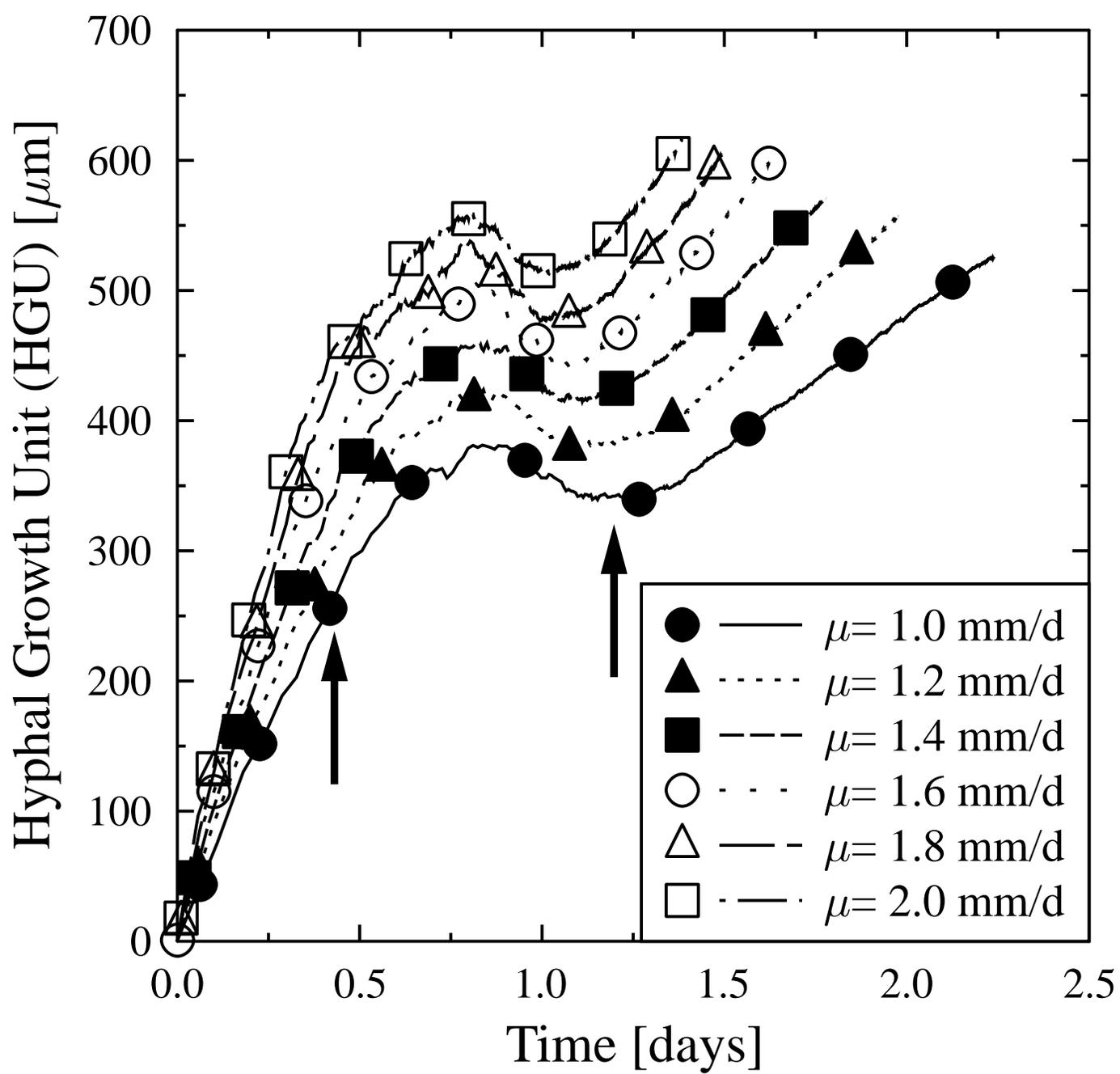



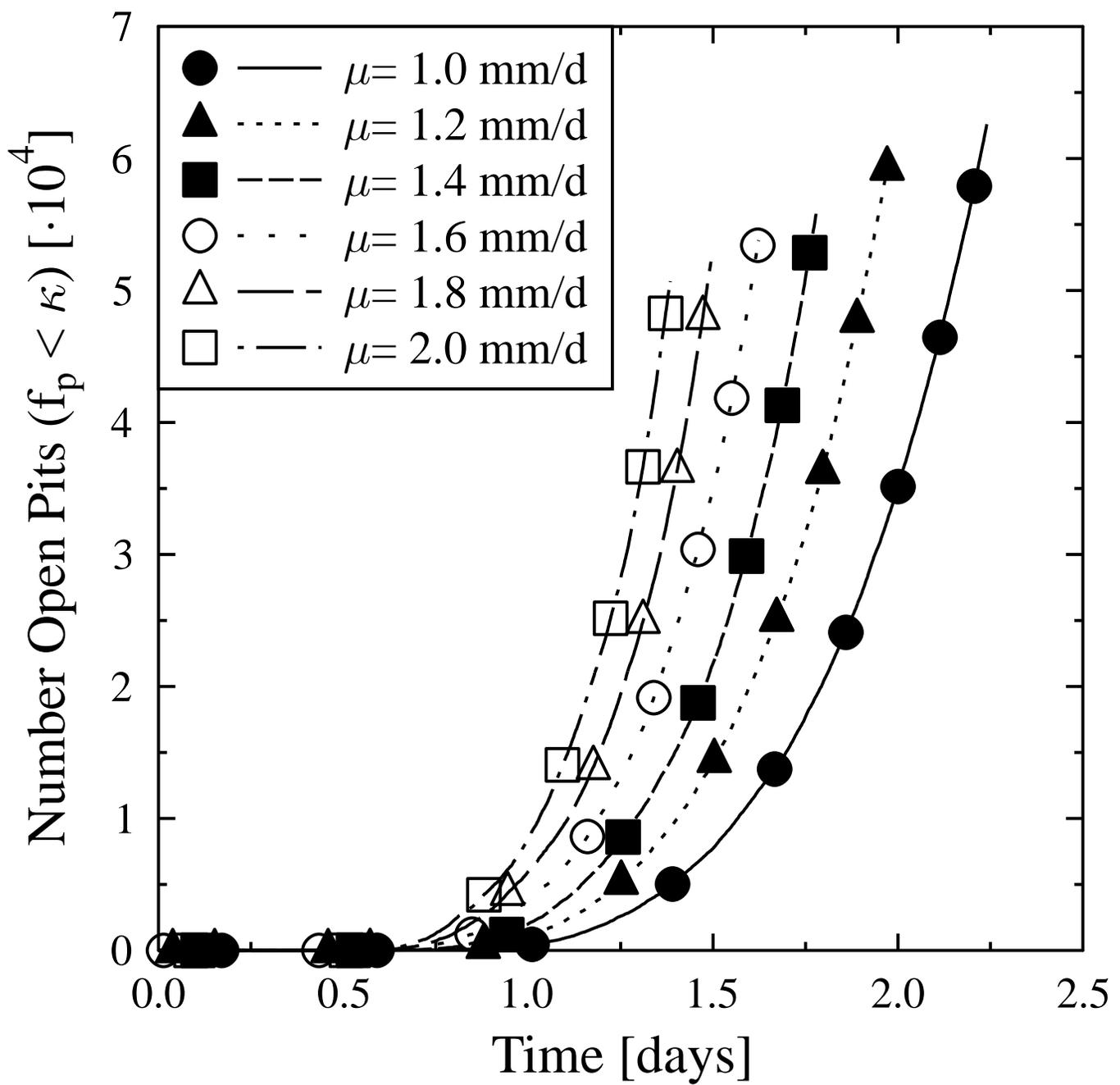



$$\varepsilon = 10^{-3}\nu$$ $$\varepsilon = 10^{-4}\nu$$

(a)
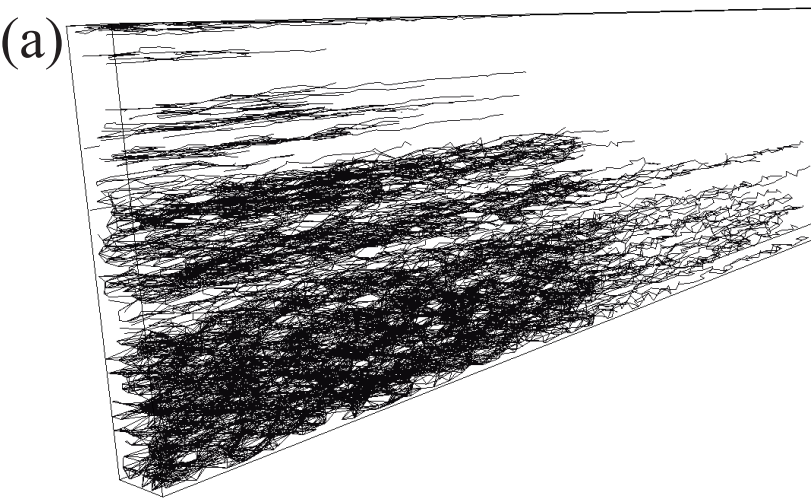

(b)
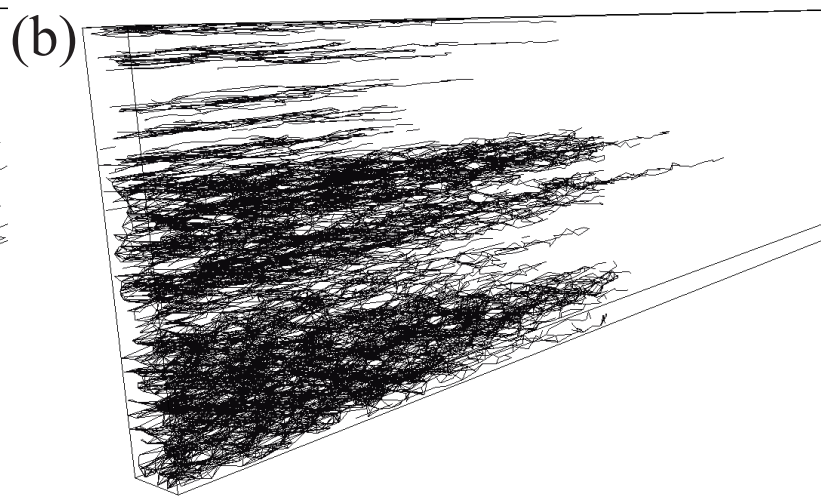

(c)
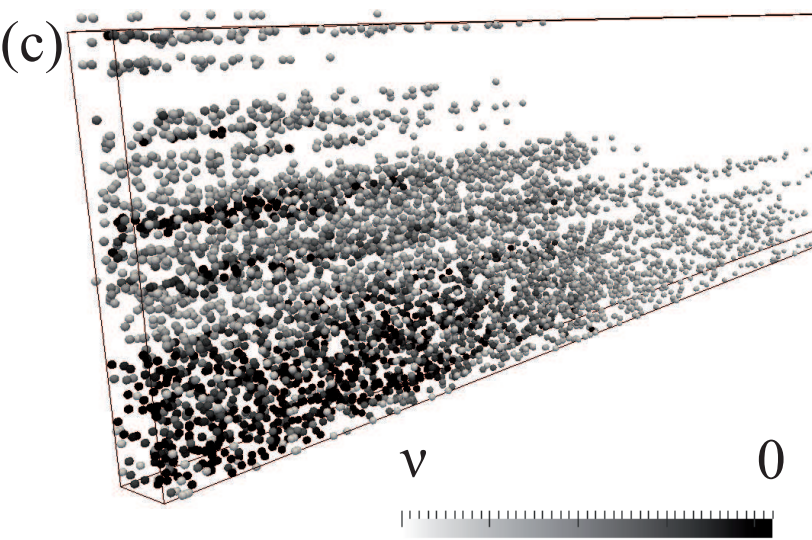

$\nu$      0

(d)
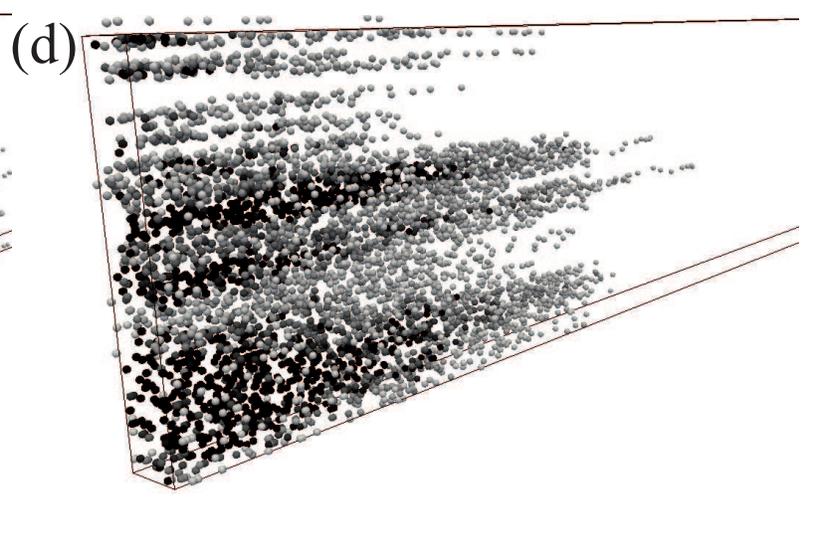

(e)
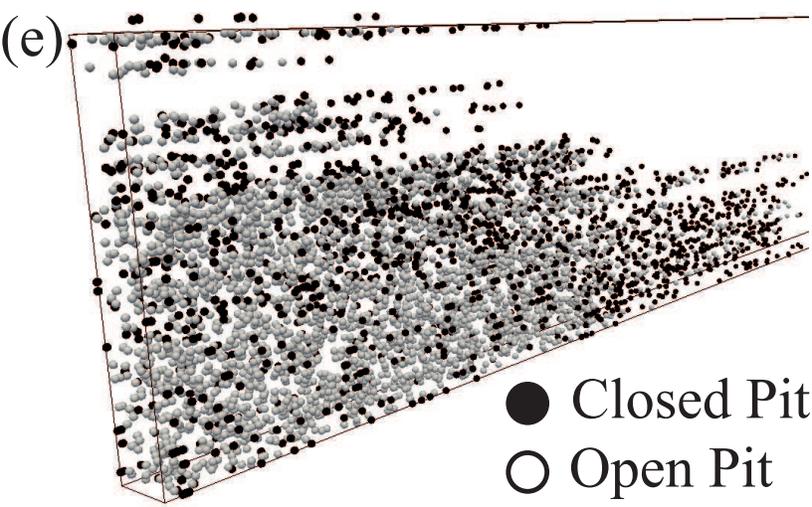

● Closed Pit
○ Open Pit

(f)
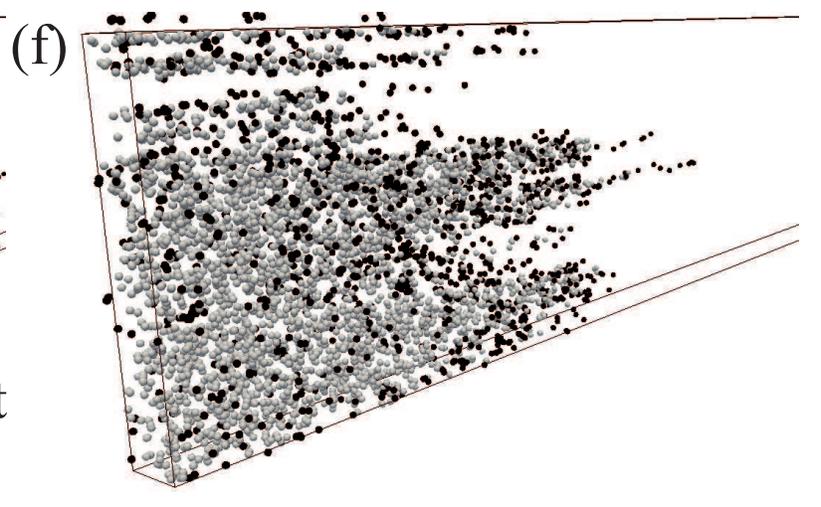